\documentclass[iop]{emulateapj}

\usepackage{amsmath}
\usepackage{float}

\slugcomment{Draft Version September 21, 2012}

\newcommand{\redmapper}{redMaPPer}

\shorttitle{Skielboe et al.}
\shortauthors{Skielboe et al.}

\begin{document}

\title{Spatial anisotropy of galaxy kinematics in Sloan Digital Sky Survey galaxy clusters}

\author{Andreas Skielboe\altaffilmark{1,2}}
\author{Rados{\l}aw Wojtak\altaffilmark{1}}
\author{Kristian Pedersen\altaffilmark{1,2}}
\author{Eduardo Rozo\altaffilmark{3,4,7}}
\author{Eli S. Rykoff\altaffilmark{5,6}}

\affil{Dark Cosmology Centre, Niels Bohr Institute, University of Copenhagen, DK-2100 Copenhagen, Denmark}
\affil{Space Science Center, Niels Bohr Institute, University of Copenhagen, DK-2100 Copenhagen, Denmark}
\affil{Department of Astronomy and Astrophysics, The University of Chicago, Chicago, IL 60637, USA}
\affil{Kavli Institute for Cosmological Physics, The University of Chicago, Chicago, IL 60637, USA}
\affil{SLAC National Accelerator Laboratory, Menlo Park, CA 94025, USA}
\affil{Lawrence Berkeley National Laboratory, Berkeley, CA 94720, USA}

\altaffiltext{7}{Einstein Fellow}

\begin{abstract}
Measurements of galaxy cluster kinematics are important in understanding the dynamical state and evolution of clusters of galaxies, as well as constraining cosmological models. While it is well established that clusters exhibit non-spherical geometries, evident in the distribution of galaxies on the sky, azimuthal variations of galaxy kinematics within clusters have yet to be observed. Here we measure the azimuthal dependence of the line-of-sight velocity dispersion profile in a stacked sample of 1743 galaxy clusters from the Sloan Digital Sky Survey (SDSS). The clusters are drawn from the SDSS DR8 redMaPPer catalog.  We find that the line-of-sight velocity dispersion of galaxies lying along the major axis of the central galaxy is larger than those that lie along the minor axis. This is the first observational detection of anisotropic kinematics of galaxies in clusters. We show that the result is consistent with predictions from numerical simulations. Furthermore we find that the degree of projected anisotropy is strongly dependent on the line-of-sight orientation of the galaxy cluster, opening new possibilities for assessing systematics in optical cluster finding.
\end{abstract}

\keywords{galaxies: clusters: general --- galaxies: kinematics and dynamics}

\section{Introduction} \label{sec:introduction}
Clusters of galaxies constitute the largest, gravitationally collapsed, structures in the universe. They offer a unique opportunity to study the formation and evolution of structure on cosmological scales. Additionally, the mass distribution of galaxy clusters can be observationally probed both dynamically and via weak lensing, making galaxy clusters ideal laboratories for studying dark matter and modifications to general relativity (e.g. \cite{2010MNRAS.406.1796R,2012arXiv1205.4679R}).

Observations and simulations clearly show that clusters exhibit triaxial rather than spherical shapes \citep{1982A&A...107..338B,2005ApJ...627..647B,2007ApJ...664..117G}. \cite{2004ApJ...606...67H} made the first detection of the flattening of galactic dark matter halos using weak lensing and \cite{2012arXiv1206.4304V} measured the azimuthal variation of the weak gravitational lensing signal around galaxies, constraining galactic dark matter halo ellipticity. Non-zero ellipticity of dark matter halos has also been confirmed in strong lensing measurements by studying the angular distribution of giant arcs around groups and clusters of galaxies \citep{2004ApJ...609...50D,2012ApJ...749...38M}.

The velocity fields of galaxies in clusters are known to feature non-trivial radial dependencies, but azimuthal variations are often overlooked. If the shape of the underlying dark matter halo is indeed multiaxial, it is natural to think that it will be reflected in an anisotropic velocity field. Indeed this has been confirmed and studied using numerical simulations \citep{1997MNRAS.290..411T,2005ApJ...629..781K,2010MNRAS.408.1818W,2012arXiv1203.5708S}. Because cluster mass estimators often assume spherically symmetric velocity fields, characterizing the degree of anisotropy in galaxy clusters is crucial in understanding its impact on cluster mass estimates. It may also provide clues to the ongoing formation of galaxy clusters.

The effect of an anisotropic velocity field should manifest itself in an azimuthal variation of the projected velocity dispersion of member galaxies. To our knowledge no attempts have been made to measure the azimuthal variation of galaxy velocity dispersions in a large sample of galaxy clusters.

The aim of this Letter is to test for azimuthal dependence of the projected velocity dispersion of cluster galaxies in a stacked sample of galaxy clusters. We use the Sloan Digital Sky Survey (SDSS) DR8 \citep{2011ApJS..193...29A} \redmapper\ cluster catalog (described below) to search for an anisotropic velocity field by measuring the velocity dispersion of cluster galaxies along the major axis of the central galaxy (CG), and galaxies along the corresponding orthogonal axis.  The results are compared with expectations from numerical simulations. We adopt a flat $\Lambda$CDM cosmology with $\Omega_m=0.3$ and $H_0 = 70$ km s$^{-1}$ Mpc$^{-1}$.

\section{Methods} \label{sec:methods}
The expected degree of the relative difference in dispersion is of the order of a few percent (see Section \ref{sec:simulations}). The strategy for detecting a signal of this magnitude is to stack a large number of galaxy clusters.

\subsection{The \redmapper\ Cluster Catalog}
Our study utilizes the SDSS DR8 red-sequence matched-filter probabilistic percolation (``\redmapper") cluster catalog (E. S. Rykoff et al. 2012, in preparation), based on the optimized richness estimator $\lambda$~\citep{2010ApJ...708..645R,2012ApJ...746..178R}.  \redmapper\ is a photometric cluster catalog that identifies galaxy clusters as overdensities of red-sequence galaxies.  The algorithm is divided into two stages: a calibration stage, where the red-sequence model is determined directly from the data, and a cluster-finding stage.  These two stages are iterated several times before a final cluster-finding run is made.

In the calibration phase, \redmapper\ empirically calibrates the color distribution (mean and scatter) of red-sequence galaxies as a function of redshift and magnitude. The calibration itself is also an iterative procedure. First, we use red galaxies from the SDSS luminous red galaxy \citep{2001AJ....122.2267E} and ``main"~\citep{2002AJ....124.1810S} spectroscopic galaxy samples (over 50\% of the sky) as training galaxies. These galaxies are used as ``seeds" to look for significant overdensities of nearby galaxies with the same color as the seed galaxy. The resulting cluster list is used to fit a full red-sequence model including zero point, tilt, and scatter. Moreover, the algorithm utilizes all colors ($u-g$, $g-r$, $r-i$, and $i-z$) simultaneously, so the ``scatter'' is characterized not by a single number but by a covariance matrix. The red sequence model is calibrated down to a luminosity threshold of  $0.2$ $L_*$ at the cluster redshift \citep[the optimal depth for cluster richness estimation][]{2012ApJ...746..178R}, thereby leveraging the bright spectroscopic galaxy sample to obtain a model of the red sequence that extends to dim magnitudes.

With our model of the red sequence at hand, we proceed to run the cluster-finding algorithm as described below. The resulting cluster catalog is then used to re-calibrate the model of the red-sequence as above, and the procedure is iterated until convergence in the red sequence model parameters is achieved. At that point, we perform one final cluster-finding run to arrive at the final \redmapper\ catalog. 

Given a red-sequence model, the cluster finding proceeds as follows. First, we consider all photometric galaxies as candidate cluster centers. We use our red-sequence model to calculate a photometric redshift for each galaxy, and evaluate the goodness of fit of our red galaxy template. Galaxies that are not a reasonable fit to the red-sequence model at any redshift are immediately discarded. For the remaining galaxies, we use this initial redshift guess and the red-sequence model to evaluate the richness ($\lambda$) and cluster likelihood using a multi-color generalization of the method in \citet{2012ApJ...746..178R}. When a significant number of red galaxies are detected, we re-estimate the cluster redshift by performing a simultaneous fit of all the high-probability cluster members to the red-sequence model. This procedure is iterated until convergence is achieved between member selection and cluster photometric redshift. The resulting list of candidate cluster centers is then rank-ordered according to likelihood, and membership probabilities are used to mask out member galaxies in the percolation step.

The final aspect of our algorithm is the selection of cluster CGs.  Specifically, if the luminosity distribution of CGs as a function of richness is known, then one can evaluate the relative probability that any given galaxy with a color consistent with the red sequence is consistent with the CG distribution, with the satellite luminosity function; or with the foreground magnitude distribution.  We also take into account a local density estimator, as CGs tend to reside in the densest parts of the cluster.  In general, however, we find that the luminosity filter dominates in most cases.  To define the filter functions, as with the red-sequence calibration we follow an iterative procedure: on our first iteration, centers are defined as the brightest high membership probability galaxy.  These assigned centers are used to characterize a ``first guess" for the luminosity distribution and local density of CGs.  We then rerun our central selection algorithm iteratively until convergence.  In the end, every galaxy in the cluster can be assigned a probability of being a CG, though typically no more than two galaxies have a significant probability of being the CG of the cluster.

\subsection{Cluster and Galaxy Selection}

To reduce an effect of stacking clusters of different sizes we choose richness cut in the catalog of \mbox{$20 < \lambda < 50$}, corresponding to the mass range $1.4 < M_{200c}/10^{14} M_\odot < 3.6 $ \citep{2012ApJ...746..178R}. When stacking clusters there is also an issue of miscentering, which we alleviate using the centering probability $P_\text{CG}$, reported in the \redmapper\ catalog, to select clusters with $P_\text{CG}>0.95$, implying we expect $>95\%$ of our galaxy clusters are properly centered. We impose no cut on redshift.

Cluster candidate members are selected from SDSS DR8. To get accurate relative velocities we use only galaxies with spectroscopic redshifts. For each CG we do an initial broad selection of member galaxies by choosing all galaxies that satisfy \mbox{$|\Delta v| < v_\text{max} = 4000$ km s$^{-1}$} and \mbox{$d < 6$ Mpc}, where $\Delta v$ is the difference in velocity between the galaxy and the CG and $d$ is their projected separation. These cuts are intentionally generous, since interlopers are included in our velocity distribution model. If a galaxy is a candidate member for several clusters, we assign the galaxy only to the nearest cluster, based on the projected distance to the CG. Finally we discard clusters with fewer than four spectroscopic candidates. To ensure that our results are insensitive to the precise selection criteria adopted, we repeat the data analysis using a large range of different criteria. We recover consistent results in all cases.

This leaves a total of 1743 clusters containing 49,524 candidate member galaxies (of which 13,565 have $d\leq2$ Mpc) that were used in the final analysis.

\subsection{Galaxy Positions Angles}

\begin{figure}
\centering
\includegraphics[width=0.9\columnwidth]{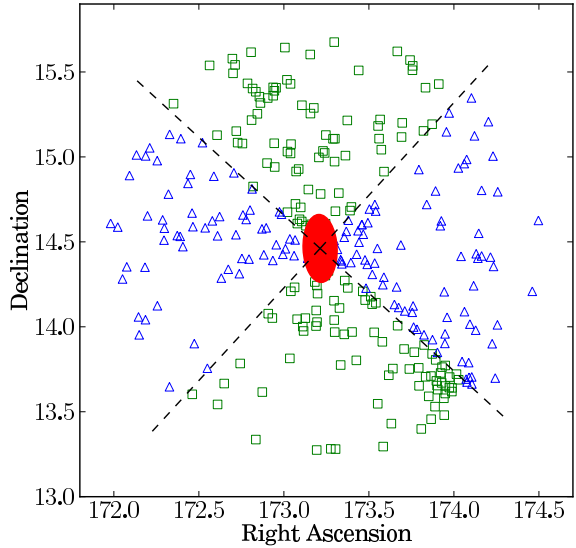}
\caption{Sky plot of a galaxy cluster with member galaxy candidates. The CG is represented by an exaggerated red ellipse. The azimuthal sections used in the analysis are indicated by dashed lines. Member candidates are marked according to their associated section, with green squares indicating galaxies closer to the CG major axis and blue triangles indicating galaxies closer to the minor axis. This plot contains interlopers as well as cluster members.}
\label{fig:cluster}
\end{figure}

CG major axis position angles are taken from the $R$-band photometric object position angles (E of N) from a de Vaucouleurs model fit provided in the SDSS DR8. To avoid ambiguity in determining CG position angles, we require that the projected axis ratio of the ellipticity for each CG is less than $0.9$. Member galaxy coordinates are used to determine the azimuthal angle $\Delta\phi$ between the galaxy and the projected CG major axis. Galaxies are put in two spatial sections according to their position angle relative to the major axis of the CG (see Figure \ref{fig:cluster}).

\subsection{Line-of-sight Galaxy-velocity Model}

Member galaxy velocities are calculated relative to the CG ($v_\text{rel} = v_\text{CG} - v_\text{gal}$). This only makes sense if the CG is at rest with respect to the halo, otherwise we might observe a systematic offset of the mean of the velocity distribution of member galaxies. If the offset is random the mean will be zero, but the dispersion will become systematically larger. We find that for all stacked velocity distributions the mean is consistent with zero, within effects from gravitational redshift (see, e.g., \cite{2011Natur.477..567W}).

The velocity distribution is modeled as a Gaussian probability distribution with mean $\mu$ and dispersion $\sigma$. Interlopers are assumed drawn from a flux-limited uniform background, which makes the background a linear decreasing function of recession velocity, given by a probability distribution with intercept $n$, and slope \mbox{$s(n) = (1/(2 v_\text{max}) - n) / v_\text{max}$} fixed by the probability distribution normalization condition. To ensure overall normalization, a parameter $p$ is introduced, which is the probability for a galaxy to be a member of a cluster. The probability for a particular relative velocity $v_\text{rel}$ in this model is given by
\begin{align}
	P(v_\text{rel}|p,\sigma,\mu,n) &= p \frac{1}{\sqrt{2\pi}\sigma} \exp \left [ {-\frac{1}{2}(v_\text{rel}-\mu)^2 / \sigma^2} \right ] \notag \\ 
	&+ (1 - p) (s (v_\text{rel} + v_\text{max})+n) \label{eq:model}.
\end{align}

Each velocity derived from SDSS includes a measurement error of the order of a few tens of km s$^{-1}$, which has the effect of slightly increasing the measured dispersion of the stacked sample of velocities. This error is small and should not affect the results, especially when studying azimuthal variation in dispersion. We estimate the likelihood for a given velocity dispersion $\sigma_v$ by a Bayesian likelihood analysis, multiplying probabilities over all velocities in the data sample,
\begin{equation}
	\mathcal{L}(p,\sigma_v,\mu,n) = \prod_i P(v_i | p,\sigma_v,\mu,n).
\end{equation}

\begin{figure}
\centering
\includegraphics[width=1.05\columnwidth]{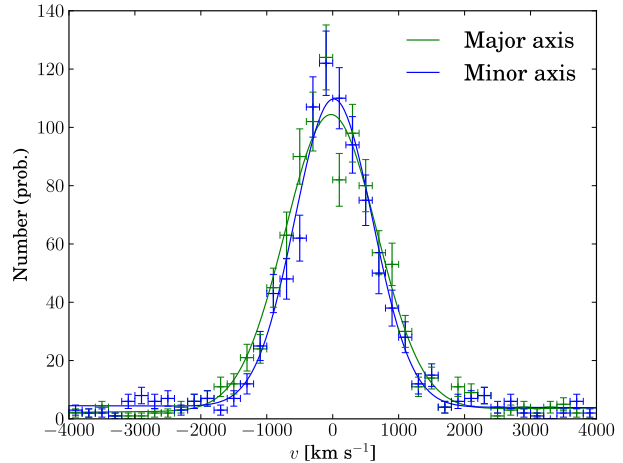}
\caption{Example velocity distribution of member galaxies along with the model from Equation (\ref{eq:model}) with highest-posterior values obtained using MultiNest (solid lines). The green crosses/line shows velocities along the major axis, and blue crosses/lines shows velocities along the minor axis.}
\label{fig:profile}
\end{figure}

To generate the posterior distribution for the parameter $\sigma_v$ we sample the likelihood over the prior parameter space, while marginalizing over all other parameters, using the method of nested sampling \citep{2008MNRAS.384..449F}. We use the publicly available software \verb+MultiNest+ \citep{2009MNRAS.398.1601F}. An example of a model fit to data for a particular radial bin is presented in Figure \ref{fig:profile}.

\section{Results} \label{sec:results}

\begin{figure}
\centering
\includegraphics[width=1.05\columnwidth]{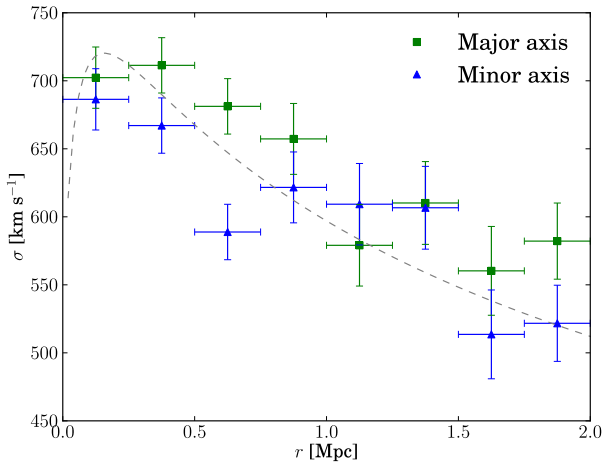}
\caption{Velocity dispersion of galaxies as a function of radius. Green squares are galaxies closer to the central galaxy major axis and blue triangles are galaxies closer to the central galaxy minor axis. The dashed gray line shows prediction from NFW.} 
\label{fig:results}
\end{figure}

Figure \ref{fig:results} shows galaxy velocity dispersion as a function of radius for two populations of cluster member galaxies; those that lie closer to the projected CG major axis and those that lie closer to the projected CG minor axis. We find a decreasing relationship between velocity dispersion and radius, consistent with the result expected for a Navarro-Frenk-White (NFW) profile \citep{1997ApJ...490..493N}, with concentration parameter $c=5.5$ \citep{2011ApJ...740..102K} and typical anisotropy of galaxy orbits in clusters \citep{2010MNRAS.408.2442W}, convolved with the mass function of the \redmapper\ clusters in the given richness range.

In Figure \ref{fig:difference} we plot the measured difference in line of sight velocity dispersion as a function of radius, divided by the mean dispersion at the given radius. The weighted difference is calculated as $\Delta\sigma_v/\langle\sigma_v\rangle = 2 (\sigma_v^\text{maj}-\sigma_v^\text{min})/(\sigma_v^\text{maj}+\sigma_v^\text{min})$. The dashed line (blue band) shows maximum likelihood ($\pm 1 \sigma$) for a constant difference in weighted dispersion between the two azimuthal sections. The posterior for the constant weighted difference is $\Delta\sigma_v/\langle\sigma_v\rangle = 0.06 \pm 0.02$, corresponding to an absolute difference of $\Delta\sigma_v = 38$ km s$^{-1} \pm 13$ km s$^{-1}$. This indicates that galaxies along the projected major axes of the CGs exhibit preferentially higher velocity dispersions compared to galaxies lying along the projected minor axes.

We test whether the observed difference could be systematic in nature by repeating our experiment, but randomly orienting the halos when stacking, and then splitting the galaxy memberships along a random axis.  As expected we do not detect any difference in the velocity dispersion of the two galaxy populations in this case.  We also verify that our conclusions are valid if we remove the radial bin with the strongest offset ($R=0.6$ Mpc), albeit at reduced significance of $ 2 \sigma$. Finally, we find that varying the cuts used to select candidate cluster members does not impact our results.

\subsection{Numerical Simulations} \label{sec:simulations}
To compare our measurement with theoretical predictions, we analyzed velocity dispersion profiles of subhalos in cluster-size halos from the Bolshoi simulation\footnote{The Bolshoi simulation is publicly available via the
MultiDark database (http://www.multidark.org; \citealt{2011arXiv1109.0003R}).} of a standard $\Lambda$CDM cosmology \citep{2011ApJ...740..102K}. We used $\sim 500$ dark matter halos with masses $M_{200c} > 0.4 \times 10^{14} M_\odot$ and subhalos with at least 30 particles corresponding to a minimum mass of $M_{200c} > 2 \times 10^9 M_\odot$. Subhalos were identified using Bound Density Maximum algorithm \citep{1997astro.ph.12217K}. For every halo we calculated projected velocity dispersion profiles as measured by a distant observer. The profiles were determined inside two equal sections along the projected major axes of the halos and along the perpendicular direction, likewise in the analysis of SDSS data. The major axes of dark matter halos were determined by diagonalization of the tensor of inertia computed using dark matter particles inside the virial spheres. We considered three fixed orientations of the line of sight with respect to the major axes of the halos, with angles $5^\circ$, $20^\circ$ and $90^\circ$, and a fourth case corresponding to a fully random angle distribution of the light cones.

\begin{figure*}
\centering
\includegraphics[width=1.5\columnwidth]{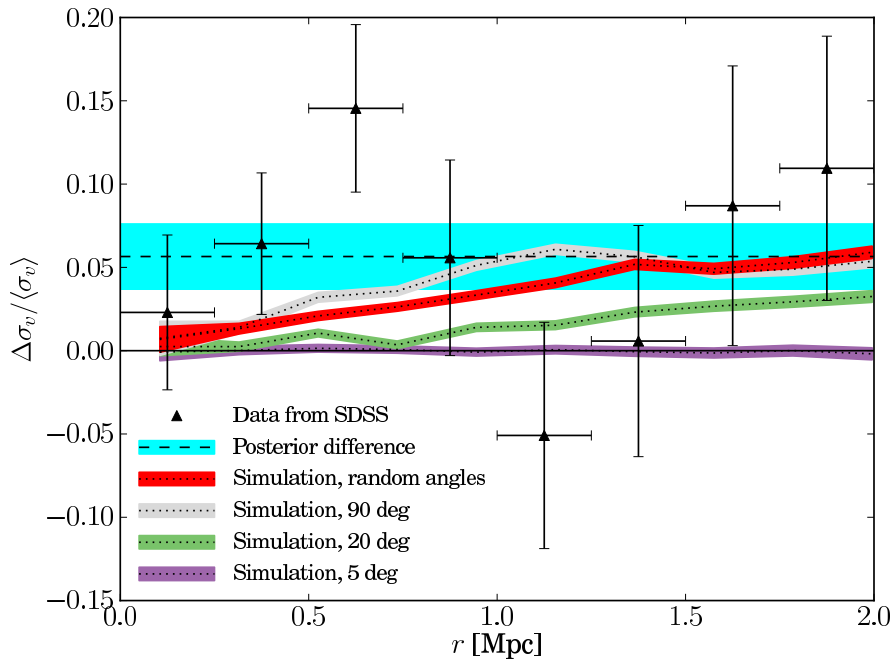}
\caption{Difference between galaxy velocity dispersion along projected major and minor axes of the CG, divided by the mean dispersion at the given radius. Positive values indicate $\sigma_v^\text{maj} > \sigma_v^\text{min}$. Triangles show results from our analysis of SDSS clusters. The dashed line (blue band) indicates the maximum likelihood for a constant offset ($\pm 1 \sigma$ errors). The dotted lines (colored bands) show results from numerical simulations ($\pm 1 \sigma$ errors from bootstrapping), with different lines corresponding to varying selection of angle between the line of sight and cluster halo major axis.} 
\label{fig:difference}
\end{figure*}

The resulting profiles are shown in Figure \ref{fig:difference}. The dotted lines show the predicted weighted dispersion difference, with the colored bands representing the error of the mean determined by bootstrapping. The mean anisotropy of the velocity dispersion increases strongly with the angle between the line of sight and the major axis. Moreover, it is positive ($\sigma_v^\text{maj} > \sigma_v^\text{min}$) at all radii. The signal vanishes for sight lines parallel to the major axis of the halo. We checked that this result does not depend on the alignment of the intermediate and minor axes of the halos. Therefore, it is a direct consequence of the fact that halos are predominantly prolate \citep{2007ApJ...664..117G}.

Even though our calculation is based on purely dark matter simulation, the signal corresponding to random orientations of major axes of dark matter halos matches the observations with $\chi_\text{sim}^2/\text{dof} = 10.4 / 8 = 1.3$ (compared to $\chi_\text{iso}^2/\text{dof} = 15.3 / 8 = 1.9$ for an isotropic case, $\sigma_v^\text{maj}=\sigma_v^\text{min}$). We checked that the relative difference of the velocity dispersions only depends weakly on the halo mass and remains the same when we restrict our calculations to the halo masses corresponding to the selected richness range.

\subsection{Systematics} \label{sec:systematics}
Miscentering of the CG has the effect of boosting the measured velocity dispersion of galaxies. Because this boost is isotropic, it will tend to dilute the signal, while simultaneously increasing the observed velocity dispersion of the cluster stack.  Furthermore, if the orientation of the ellipticity of the false centers is random this will likewise act to dilute the signal. We have attempted to minimize these systematics by requiring that all clusters be properly centered ($P_\text{CG}>0.95$), but strictly speaking our measurement can be thought of as a lower limit.

While CGs and host halos tend to be preferentially aligned in observed galaxy clusters, there is also a significant random component associated with the angular separation of their major axes \citep{2010MNRAS.405.2023N}. This will tend dilute the observed signal relative to our predictions from numerical simulations. Thus, similarly in this case, one can think of our results as providing a lower limit to the anisotropic signal relative to the halo's major axis.

An additional concern is that because halo triaxiality depends on halo mass, the relative satellite-weighted contribution of a single halo in a stack will be different along the major and minor axes.  In other words, the effective halo mass distribution for stacks along the major and minor axes will be different, and this could lead to an artificial anisotropy signal.  We have estimated the level of this effect by use of a toy model, and found that, even when triaxiality scales strongly with halo mass, the effect is negligible.

\section{Discussion and conclusion}
We present the first observational detection of anisotropic kinematics of galaxies in clusters of galaxies. We have measured line-of-sight velocity dispersions of member galaxies along the major and minor axes of the CG, in a large sample of stacked clusters from the SDSS. The projected CG position angle is used to separate galaxies into two section depending on their projected proximity to either axes. Galaxies closer to the projected CG major axis are found to have a preferentially larger velocity dispersion than those that are closer to the minor axis, with an average difference in dispersion of $\Delta\sigma_v = 38$ km s$^{-1} \pm 13$ km s$^{-1}$, corresponding to a relative difference of $\Delta\sigma_v / \langle\sigma_v\rangle = 6$\% $\pm$ 2\%. These results are consistent with results from cosmological simulations, with $\chi_\text{sim}^2/\text{dof} = 10.4 / 8 = 1.3$.

The presence of this effect is a signature of a prolate velocity ellipsoid in galaxy clusters. Keeping in mind that velocity ellipsoid is aligned with the halo shape ellipsoid \citep{2005ApJ...629..781K,1997MNRAS.290..411T,2010MNRAS.408.1818W} and the halos tend to be oriented with the cosmic web \citep{2006MNRAS.370.1422A,2007MNRAS.375..184B}, the effect is likely a remnant of an anisotropic halo formation (\citealt{2012JCAP...07..042S}; R. Wojtak et al. 2012, in preparation). This raises an important question as to what extent cluster dynamics is influenced by the infall process and cluster location in the cosmic web. The result can also be interpreted in light of studies indicating preferential alignment between the CG and its host halo \citep{1982A&A...107..338B, 2008MNRAS.390.1562H, 2010MNRAS.405.2023N}, strengthening the case for anisotropic cluster formation.

We furthermore find a clear correlation between the degree of projected anisotropy and the line-of-sight orientation of cluster halos in the simulations. This result opens exciting new possibilities for assessing the degree of random alignment of galaxy clusters. Specifically it might be applied to a sample of optically selected clusters to evaluate the level of selection bias.

\acknowledgments
We warmly thank the referee, Andrea Biviano, for his comments and help in improving the manuscript. We thank Martin Sparre, Surhud More and Matthew Becker for helpful comments. Dark Cosmology Centre is funded by the Danish National Research Foundation. Kristian Pedersen acknowledges support from Instrument Center for Danish Astrophysics. Eduardo Rozo is funded by NASA through the Einstein Fellowship Program, grant PF9-00068. This work was supported in part by the U.S. Department of Energy contract to SLAC No. DE-AC02-76SF00515.

\end{document}